\documentclass{optica-article}

\journal{opticajournal} 

\articletype{Research Article}

\usepackage{lineno}

\usepackage{amsmath}
\usepackage{float}
\RequirePackage{graphicx}
\usepackage{textgreek}
\usepackage{siunitx}
\usepackage{xfrac}
\usepackage[shortlabels]{enumitem}
\usepackage{hyperref}

\begin{document}

\title{Conic Surfaces and Transformations for X-Ray Beamline Optics Modeling}

\author{Manuel Sanchez del Rio,\authormark{1,*} and Kenneth Goldberg\authormark{2}}

\address{\authormark{1}European Synchrotron Radiation Facility, 71 Avenue des Martyrs, 38000 Grenoble, France\\
\authormark{2}Lawrence Berkeley National Laboratory, Berkeley CA, USA}

\email{\authormark{*}srio@esrf.eu} 


\begin{abstract*} 
Optical surfaces represented by second-degree polynomials (quadratic or conics) are ubiquitous in optics. We revisit the equations of the conic shapes in the context of grazing incidence optics, gathering together the curves commonly used in x-ray instruments and synchrotron beamlines. We present the equations for paraboloids, ellipsoids, and hyperboloids in a common and consistent notation. We develop the transformations from centered systems that are commonly used to describe conics and their axes of symmetry, to local coordinate systems centered on the off-axis mirror surfaces. The equations presented are directly applicable to ray tracing, fabrication, and metrology calculations. They can also be used to study misalignments, movement tolerances, and aberrations of optical surfaces.
\end{abstract*}


\section{Introduction}\label{sec:intro}

Conic surfaces have various properties and applications in geometry, physics, engineering, and computer graphics.
They play a crucial role in the field of optics, where they are used to describe the shapes of optical surfaces such as lenses and mirrors, for applications in telescopes, microscopes, cameras, and laser systems. These surfaces are formed by rotation of a conic section (parabola, ellipse, hyperbola, circle) around an axis of symmetry. 

As mirrors, these ideal shapes are used in the following ways. Ellipsoidal surfaces focus point-to-point\cite{Goldberg2022b}. Paraboloidal surfaces focus collimated light, or collimate diverging light \cite{Goldberg2022a}. Hyperboloidal surfaces change the apparent focal distances of diverging or converging light\cite{Goldberg2023}.

This paper aims to provide an overview of the mathematical equations of the conic surfaces in optics in general non-centered systems, where the light beam is not traveling along a surface symmetry axis. This is the case of grazing incidence optics, essential for x-ray instruments and synchrotron radiation beamlines\cite{Peatman1997}. The equations of these surfaces are expressed as a function of the design conjugate distances and central angle, or \emph{design parameters}: this is typically the distance source-component $p$, the distance component-image plane $q$, and the grazing incidence angle at the optical surface $\theta$. For high-performance and wavefront-preserving x-ray, extreme-ultraviolet, and other short-wavelength applications, optical surface shapes are now routinely measured in fractions of a nanometer, even when they extend hundreds of millimeters or over a meter in length. This puts great demands on the precision of their mathematical description and analysis. 

Coddington's Equations provide a first approximation of the focusing properties of optical elements based on their local curvature \cite{Kingslake1994}. With paraxial focal length $f$, and glancing angle of incidence $\theta$, we have the following pair of relations:
\begin{equation}
\label{Eq:RmRs}
R_s = 2f \sin\theta\;\;\;\mathrm{and}\;\;\;R_m = \dfrac{2 f }{\sin\theta}.
\end{equation}
Here, at the central point of intersection, $R_m$ and $R_s$ are the meridional (tangential) and sagittal radii of curvature, respectively. As surface descriptions increase in complexity, these expressions can provide a useful means of validation.


Fermat's principle, or the principle of least time, states that the path taken by a ray of light between two points is the path that can be traversed in the least time. This principle can be used to derive the laws of reflection and refraction \cite{Hecht}. The equations of conic mirrors may be directly deduced from the Fresnel principle, like for parabolic \cite{Goldberg-parabola} and elliptical \cite{Goldberg-ellipse} mirrors. 

In this paper, we provide a mathematical analysis of the equations governing conic surfaces and derive the parameters for various types of conic mirrors  based on manufacturing specifications.
Our goal is to provide a comprehensive and consistent mathematical overview of these surfaces, which is valuable for applications in metrology and computational ray tracing.

\section{The conic coefficients}\label{sec:conic_coefficients}

A generic optical surface can be expressed by its implicit equation 
\begin{equation}
\label{eq:generic_surface_vector}
F(x,y,z)=0.
\end{equation}

In this paper, we concentrate on the particular case that $F$ is a second-degree polynomial (quadratic). This case covers many primitive optics of high interest in optics, like planes, spheres, ellipsoids, paraboloids, hyperboloids, and their corresponding ``sagittally-plane" versions (cylinders of parabolic, elliptical, or hyperbolic section).
These surfaces  belong to the generic family of conics, which are expressed as
\begin{equation}
\label{eq:conic}
\begin{split}
 F(x,y,z) =~& c_{xx} x^2
 + c_{yy} y^2
 + c_{zz} z^2 \\
 & + c_{xy} x y
 + c_{yz} y z 
 + c_{xz} x z \\ 
 & + c_x x + c_y y + c_z z + c_0 = 0.
\end{split}
\end{equation}

The optical element surface, or explicit function $z(x,y)$, is obtained first by rearranging the terms in $z$: $F(x,y,z)=A z^2 + B z + C$ with $A=c_{zz}; B=c_{yz} y + c_{xz} x + c_z; C=c_{xy} x y + c_x x + c_y y +c_0$, and solving the second-degree equation for each coordinate $(x,y)$ in the basal plane of the mirror. 

A central problem in ray tracing is to calculate the intersection of the optical surface [Eq.~(\ref{eq:generic_surface_vector})] with a ray.
The ray can be defined by a point $(x_0,y_0,z_0)$ and a director vector $(v_x,v_y,v_z)$. The ray propagates in vacuum along a straight line of coordinates $(x,y,z)$ expressed by the parametric equations
\begin{equation}
\label{eq:ray}
\begin{split}
x=x_0 & + t v_x \\
y=y_0 & + t v_y \\
z=z_0 & + t v_z,
\end{split}
\end{equation}
where $t$ is the free parameter ($t$ is the time if $(v_x,v_y,v_z)$ is the velocity, or the optical path if it is the unitary direction vector).
Introducing Eq.~(\ref{eq:ray}) into (\ref{eq:conic}) we obtain a second degree equation in $t$ whose solutions fully define the two, one, or zero intersection points.

For practical purposes, especially for non-centered optical systems using grazing incidence optics, Eq.~(\ref{eq:conic}) is expressed in a \emph{mirror-centered} reference system local to the mirror, with the coordinate origin at the center of the mirror. In the reference system in this paper (as in the SHADOW ray-tracing code \cite{shadow}), the $x$ axis goes along the sagittal (transverse) direction, the $y$ axis goes along the meridional direction (tangential, pointing close to the direction of the beam), and $z$ is the surface height measured orthogonal to the $xy$-plane. In this coordinate-system definition, the $xy$ plane is tangent to the surface at the central point.

We will obtain the $c$ coefficients of Eq.~(\ref{eq:conic}) for the different conic optical elements or mirrors placed in a non-centered system, like a typical synchrotron radiation beamline.

\section{Transformation of the conic coefficients}\label{sec:conic-coefficients}

The generic conic in Eq.~(\ref{eq:conic}) can be written in matrix form
\begin{equation}\label{eq:conicmatrix}
F(\textbf{X})=\textbf{X}^T C_M \textbf{X} + \textbf{C}_V^T \textbf{X} + c_0 = 0, 
\end{equation}
with 
\begin{equation}\label{eq:matrices}
C_M = \begin{bmatrix} 
c_{xx} & \frac{c_{xy}}{2} & \frac{c_{xz}}{2} \\
\frac{c_{xy}}{2} & c_{yy} & \frac{c_{yz}}{2} \\
\frac{c_{xz}}{2} & \frac{c_{yz}}{2} & {c_{zz}}
\end{bmatrix}; 
\textbf{C}_V = 
\begin{bmatrix}
    c_x\\
    c_y\\
    c_z
\end{bmatrix}
; \textbf{X}=
\begin{bmatrix}
x\\
y\\
z
\end{bmatrix}
\end{equation}
and the $\text{T}$ is the transpose operator. We want to modify the conic surface by a sequential rotation and translation. The transformed conic has the same shape, so the transformation can be seen as a change of the reference system. We want to calculate the transformed coefficients $c'_i$ as a function of the original coefficients $c_i$ in Eq.~(\ref{eq:conic}), a rotation matrix $R_M$ and a translation vector $\textbf{t}$. We follow the text in Ref. \cite{penelope2018}. 

The matrix $R_M$ describes a 3D rotation. A rotation around an arbitrary axis can be decomposed into sequential rotations, for example, rotation about the Cartesian axes, or by using the Euler angles
(in \cite{penelope2018} the three Euler angles $\Omega,\Theta,\Phi$ are sequential rotations around the $Z$, $X$ and, again, $Z$ axes).  
Given a conic $F(\textbf{X}) = 0$, we can generate a new surface by applying a rotation $R_M(\Omega,\Theta,\Phi)$ and a translation $T(\textbf{t})$  (in this
order). The implicit equation of the transformed surface $G(\textbf{X})$ in its new mirror-centered coordinate system is
\begin{equation}
G(\textbf{X}) = F(\textbf{X}') = F[R_M^{-1}(\Omega,\Theta,\Phi) T^{-1} (\textbf{t}) \textbf{X} ]= 0,
\end{equation}
which states that $G(\textbf{X})$ equals the value of the original function at the point 
$\textbf{X}' =  R_M^{-1} (\Omega, \Theta, \Phi) T^{-1} (\textbf{t}) \textbf{X}$
that transforms into $\textbf{X}$ [i.e., $\textbf{X} = T(\textbf{t}) R_M(\Omega, \Theta, \Phi)\textbf{X}'$].
Therefore, the equation for the rotated-shifted conic is, from Eq. (\ref{eq:conicmatrix}),
\begin{equation}
    (\textbf{X} - \textbf{t})^T R_M C_M R_M^T (\textbf{X}-\textbf{t}) + (R_M \textbf{C}_V)^T(\textbf{X}-\textbf{t}) + c_0 = 0.
\end{equation}
This can be reduced to the generic form
\begin{equation}
    \textbf{X}^T \hat{C}_M \textbf{X} + \hat{\textbf{C}}_V^T \textbf{X} + \hat{c_0} = 0, 
\end{equation}
with
\begin{equation}\label{eq:transformedcoefficients}
    \begin{split}
    \hat{C}_M=R_M C_M R_M^T\\
    \hat{\textbf{C}}_V = R_M \textbf{C}_V - 2 \hat{C}_M \textbf{t}\\
    \hat{c}_0 = c_0 + \textbf{t}^T(\hat{C}_M\textbf{t}-R_M \textbf{C}_V),
    \end{split}
\end{equation}
which fully defines the new coefficients $c'_i$. 

\section{Conics in a centered system with origin in a focus}

As an example of the application of the equations~(\ref{eq:transformedcoefficients}), we translate the centered conics to place the coordinate origin at one focus.
This geometry arises when studying a combination of two or more optical elements aligned on a central axis, as with a Wolter telescope \cite{MangusUnderwood, Shenghao_Chen2016}, or a Schwarzschild objective \cite{schwarzschild1905,Sironi2017}. We show that computation with the matrix formalism matches direct variable substitution in this simple case.

\begin{figure}
\centering
a)~~~~~~~~~~~~~~~~~~~~~~~~~~~~~~~~~~~~~~~~~~~~~~~~~~~~~~~~~~~~~~~~~~~~~~~~~~~~~~~~~~~~~~~~~~~~~~~~~~~~~~~~~~~~~~~~~~~~~\\
\includegraphics[width=0.85\textwidth]{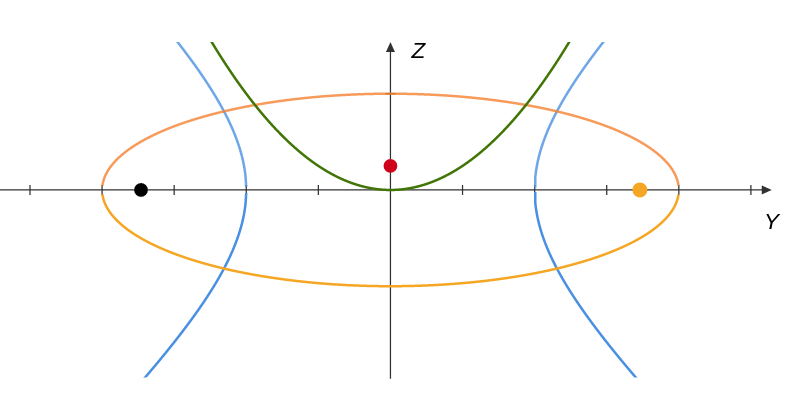}
b)~~~~~~~~~~~~~~~~~~~~~~~~~~~~~~~~~~~~~~~~~~~~~~~~~~~~~~~~~~~~~~~~~~~~~~~~~~~~~~~~~~~~~~~~~~~~~~~~~~~~~~~~~~~~~~~~~~~~~\\
\includegraphics[width=0.85\textwidth]{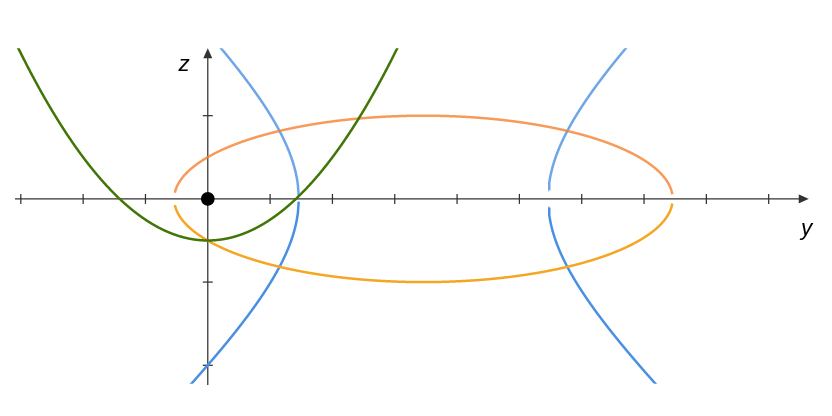}
    \caption{a) Ellipse (orange), hyperbola (blue) and parabola (green) in the centered system, showing in black one focus of the ellipse and hyperbola, and in red the focus of the parabola. b) the same conics expressed in a reference system with origin in their respective foci (see text). }\label{fig:centered}
\end{figure}

The equations of the two-dimensional, centered parabola, ellipse, and hyperbola, in the plane $(Y,Z)$ serve as the generating curves for the axial-symmetric, three-dimensional shapes in $(X,Y,Z)$. In this centered system, the coordinates are capitalized. 
Fig.~\ref{fig:centered}a displays the three conics in this centered system. Table~\ref{tab:centered_table} shows $(X,Y,Z)$, the coefficient matrix $C_M$, the vector $\textbf{C}_V$, and the independent term $c_0$.

\begin{table}[]
\centering
\begin{tabular}{lccc}
    \hline
    & parabola  & ellipse  &   hyperbola \\
    \hline
    Equation &
    $Y^2 =  4 a_p Z$ &
    $\frac{Y^2}{a_e^2} + \frac{Z^2} {b_e^2} = 1$ &
    $\frac{Y^2}{a_h^2} - \frac{Z^2} {b_h^2} =1$\\
\hline
    $C_M$ & 
$\begin{bmatrix} 
0 & 0 & 0 \\
0 & 1 & 0 \\
0 & 0 & 0 
\end{bmatrix}$ &
$\begin{bmatrix} 
0 & 0 & 0 \\
0 & \frac{1}{a_e^2} & 0 \\
0 & 0 & \frac{1}{b_e^2} 
\end{bmatrix}$ &
$\begin{bmatrix} 
0 & 0 & 0 \\
0 & \frac{1}{a_h^2} & 0 \\
0 & 0 & -\frac{1}{b_h^2}
\end{bmatrix}$ \\
\hline
$\textbf{C}_V$ &
$\begin{bmatrix}
0\\
0\\
-4a_p
\end{bmatrix}$ &
$\begin{bmatrix}
0\\
0\\
0
\end{bmatrix}$ &
$\begin{bmatrix}
0\\
0\\
0
\end{bmatrix}$ \\
\hline
$c_0$ &
0 & -1 & -1 \\
\hline
\end{tabular}
\caption{Centered conics in the 2D $(Y,Z)$ plane. These are the generating curves of the axial-symmetric, three-dimensional shapes.}
\label{tab:centered_table}
\end{table}

To bring the center to the focal position, we define the translator vector as follows.
\begin{itemize}
    \item For the parabola: $\textbf{t}=[0,0,-a_p]^T$.
      \item For the ellipse: $\textbf{t}=[0,c_e,0]^T$.
    \item For the hyperbola: $\textbf{t}=[0,c_h,0]^T$.
\end{itemize}
Here, $a_p$ is the focal distance (Eq.~\ref{eq:centeredconics}a)
$c_e$ and $c_h$ are
the distances from the origin to the focus of the ellipse and the hyperbola, respectively.

We are interested in the simple case of a pure translation, therefore the equations of the conics in a reference system $(x,y,z)$ centered at the focus,  can be obtained directly, by substitution:
\begin{subequations}\label{eq:centeredconics}
    \begin{align}
    \mathrm{parabola:~} y^2 =  4 a_p (z+a_p), \\
    \mathrm{ellipse:~} \frac{(y-c_e)^2}{a_e^2} + \frac{z^2} {b_e^2} = 1,\\
    \mathrm{hyperbola:~} \frac{(y-c_h)^2}{a_h^2} - \frac{z^2} {b_h^2} =1.
    \end{align}
\end{subequations}
Nonetheless, as an exercise, we want to obtain them by applying the Eqs.~(\ref{eq:transformedcoefficients}). We observe that there is no rotation, therefore $R_M$ and $R_M^T$ are the identity matrix. In consequence, $\hat{C}_M=C_M$ so the coefficients of second-order do not change. Moreover, $\hat{\textbf{C}}_V=\textbf{C}_V-2 C_M \textbf{t}$, and $\hat{c}_0=c_0+\textbf{t}^T(C_M\textbf{t}-\textbf{C}_V)$. A summary of the coefficients is in Table~\ref{tab:centered_table_shifted}. The conic equations obtained in this way (Table~\ref{tab:centered_table_shifted}) are equivalent to those directly deduced [Eq.~(\ref{eq:centeredconics})].

\begin{table}[]
\centering
\begin{tabular}{lccc}
    \hline
    & parabola  & ellipse  &   hyperbola \\
\hline
    $\hat{C}_M$ & 
$\begin{bmatrix} 
0 & 0 & 0 \\
0 & 1 & 0 \\
0 & 0 & 0 
\end{bmatrix}$ &
$\begin{bmatrix} 
0 & 0 & 0 \\
0 & \frac{1}{a_e^2} & 0 \\
0 & 0 & \frac{1}{b_e^2} 
\end{bmatrix}$ &
$\begin{bmatrix} 
0 & 0 & 0 \\
0 & \frac{1}{a_h^2} & 0 \\
0 & 0 & -\frac{1}{b_h^2}
\end{bmatrix}$ \\
\hline
$\hat{\textbf{C}}_V$ &
$\begin{bmatrix}
0\\
0\\
-4a_p
\end{bmatrix}$ &
$\begin{bmatrix}
0\\
-\frac{2c_e}{a_e^2}\\
0
\end{bmatrix}$ &
$\begin{bmatrix}
0\\
-\frac{2c_h}{a_h^2}\\
0
\end{bmatrix}$ \\
\hline
$
\hat{c}_0$ &
$-4 a_p^2$ & $-1 + \frac{c_e^2}{a_e^2}$ & $-1 + \frac{c_h^2}{a_h^2}$ \\
\hline
    Equation &
    $y^2 -  4 a_p z -4a_p^2$  = 0;&
    $\frac{y^2}{a_e^2} + \frac{z^2} {b_e^2} - 2\frac{c_e}{a_e^2}y - 1 + \frac{c_e^2}{a_e^2} =0$; &
    $\frac{y^2}{a_h^2} - \frac{z^2} {b_h^2} - 2\frac{c_h}{a_h^2}y - 1 + \frac{c_h^2}{a_h^2} =0$;\\
\end{tabular}
\caption{Conics in the 2D $(y,z)$ plane, centered on one focus. They are obtained from the centered conics of Table~\ref{tab:centered_table} transformed using Eqs.~(\ref{eq:transformedcoefficients}).}
\label{tab:centered_table_shifted}
\end{table}


\section{General analytic descriptions of paraboloids, ellipsoids, and hyperboloids in surface-centered coordinates}\label{sec:SectionConics}

For studying mirrors in synchrotron beamlines, for computing the surface height profile $z(x,y)$, for surface fabrication and metrology, and for ray-tracing calculations, it is useful to express Eq.~(\ref{eq:conic}) in a system local to the optical element, with zero coordinate and zero slope at its center. We refer to this as the \emph{local} coordinate system; it is often called the \emph{mirror-centered} system. It is often desirable to describe the mirror height profile $z(x,y)$---and the coefficients of Eq.~(\ref{eq:conic})---as a function of the design parameters: the conjugate distances $p$ and $q$ (distances source-mirror and mirror-image, respectively), and the glancing angle of incidence $\theta$, measured at the central-ray's intersection point.

We apply rotation and translation to the conic sections to bring them from their centered system to the local coordinate system.
In this way, we obtain polynomial expressions of the transformed surfaces, as $z_i(x_i,y_i)$, a format that simplifies the study of arbitrary misalignments and surface errors.

Starting with the centered conic, we follow these steps to compute the conic coefficients in the local reference: 
\begin{enumerate}
    \item Write the conic equation in the centered reference system and compute the list of the 10 coefficients $c_i$. In a centered system, the rotational symmetry around the $z$-axis implies some coefficients are zero:  $c_{xy}=c_{yz}=c_{xz}=0$.
    The coordinates in this centered system are capitalized: $(X,Y,Z)$. Using these coefficients, construct the $C_M$ matrix and $\textbf{C}_V$ vector [Eq.~(\ref{eq:matrices}), Table~\ref{tab:centered_table}].
    \item Express the parameters that appear in the coefficients (typically the semi-axes, linear eccentricity, etc.) as a function of the ``design parameters" (conjugate distances: source-mirror distance $p$, mirror-focus distance $q$, and glancing incidence angle $\theta$ )
    \item Compute the coordinates of the center of the mirror $\textbf{X}_c=(X_c,Y_c,Z_c)$ (this point will be the origin in local coordinate system). 
    \item Compute the
    normal versor $\textbf{n}$ at $\textbf{X}_c$, using the expression $\textbf{N}=-{\bigtriangledown}F$; thus $\textbf{n} = \textbf{N}/|\textbf{N}| =  (n_x,n_y,n_z)$.
    \item Compute the roto-translation matrix and vector, to implement the sequential rotation of $\Theta$ around the $X$ axis (to bring $\textbf{N}$ to the new $\textbf{z}$ axis) and a translation $\textbf{t}$, to bring $\textbf{X}_c$ to the origin of the new system. 
    The rotation around $X$ of angle $\Theta$, has a matrix
    \begin{equation}\label{eq:rotX}
    R_{M,X}(\Theta) = \begin{bmatrix} 
    1 & 0 & 0 \\
    0 & \cos\Theta & -\sin\Theta \\
    0 & \sin\Theta & \cos\Theta
    \end{bmatrix},
    \end{equation}
    and the translation vector is $\textbf{t}=-R_M \textbf{X}_c$. Compute $\Theta$ as a function of the known parameters.
    \item Apply the Eqs.~(\ref{eq:transformedcoefficients}) to get the new set of coefficients $c'_i$. 
\end{enumerate}

\subsection{Paraboloid}
\begin{figure}
\centering
\includegraphics[width=0.85\textwidth]{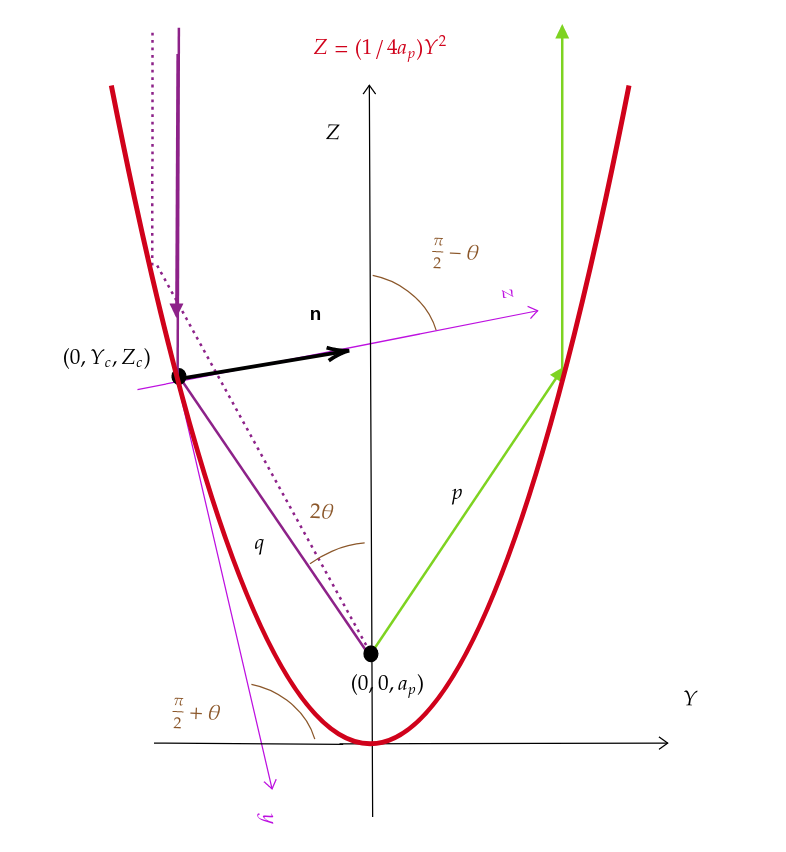}
    \caption{The parabola in a centered reference system $(X,Y,Z)$. The local reference system is $(x,y,z)$}\label{fig:parabola}
\end{figure}

The meridional section in the plane $YZ$ has equation  (Fig.~\ref{fig:parabola}, Table~\ref{tab:centered_table})
\begin{equation}
    \label{eq:parabola}
    Y^2 = 4 a_p Z, 
\end{equation}
with focus at $(0,a_p,0)$ and vertex at $(0,0,0)$. The distance from the focus to the parabola directrix is
$2a_p$. The non-zero coefficients of the revolution paraboloid (obtained by replacing $Y^2\xrightarrow{} X^2+Y^2$) are $c_{yy}=c_{xx}=1,c_z=-4a_p$.
Two configurations are possible: (i) focusing incident rays, collimated parallel to the axis of symmetry, and (ii) collimating rays emanating from the focus.

i) Focusing configuration (type IIIA in \cite{Goldberg-parabola}, violet path in Fig.~\ref{fig:parabola}), with
\begin{itemize}
    \item The design parameters are $p=\infty,~q,~\theta$.
    \item The center abscissa is $Y_c=-q \sin(2\theta)=-2q\sin\theta \cos\theta$.
    \item The parameter $a_p$ can be obtained from the slope at $\textbf{X}_c$ which is $m=d(z(Y_c))/dY=Y_c/(2a_p)$. Therefore,  $a_p=Y_c/(2 m)=Y_c/(2\tan(\pi/2 + \theta))=- Y_c \tan(\theta)/2=q \sin^2\theta$.
    \item The center ordinate is $Z_c=Y_C^2/(4 a_p)=q \cos^2\theta$.
    \item The (non-normalized) normal vector is aligned with $\textbf{N}=(0,-2Y_c,4 a_p)$.
    \item The normalized normal is $\textbf{n}=\textbf{N}/|\textbf{N}|=(0,\cos\theta,\sin\theta)$.
    \item The angle from $\textbf{n}$ to $\textbf{Z}$ is $\Theta=\pi/2-\theta$ .
    \item The non-zero $c'$ coefficients are 
\begin{equation}
    \label{eq:coeffs-paraboloid3a}
    \begin{alignedat}{2}
    c'_{xx}&=1, \\
    c'_{yy}&=\sin^2\theta &&=n_z^2,\\
    c'_{zz}&=\cos^2\theta &&=n_y^2,\\
    c'_{yz}&=2 \sin\theta \cos\theta &&=2 n_y n_z,\\
    c'_z&= -4 q \sin\theta &&=-4 q n_z.
    \end{alignedat}
\end{equation}
\end{itemize} 

ii) Collimating mirror (type IIIB in \cite{Goldberg-parabola}, green path in Fig.~\ref{fig:parabola}), with
\begin{itemize}
    \item The design parameters are $p,~q=\infty,~\theta$
    \item  The center absciss is $Y_c=p \sin(2\theta)=2p\sin\theta \cos\theta$.
    \item The parameter $a_p$ is $a_p=p \sin^2\theta$.
    \item The center ordinate is $Z_c=Y_C^2/(4 a_p)=p \cos^2\theta$.
    \item The (non-normalized) normal is aligned with $\textbf{N}=(0,-2Y_c,4 a_p)$.
    \item The unit normal vector is $\textbf{n}=\textbf{N}/|\textbf{N}|=(0,-\cos\theta,\sin\theta)$
        \item The angle from $\textbf{n}$ to $\textbf{Z}$ is $\Theta=-(\pi/2-\theta)$ .
    \item The non-zero $c'$ coefficients are 
\begin{equation}
    \label{eq:coeffs-paraboloid3a}
    \begin{alignedat}{2}
    c'_{xx}&=1, \\
    c'_{yy}&=\sin^2\theta &&=n_z^2,\\
    c'_{zz}&=\cos^2\theta &&=n_y^2,\\
    c'_{yz}&=-2 \sin\theta \cos\theta&& =2 n_y n_z, \\
    c'_z&= -4 p \sin\theta &&=-4 p n_z.
    \end{alignedat}
\end{equation}
\end{itemize} 

\subsection{Ellipsoid}\label{sec:ellipse}

\begin{figure}
    \centering    \includegraphics[width=0.85\textwidth]{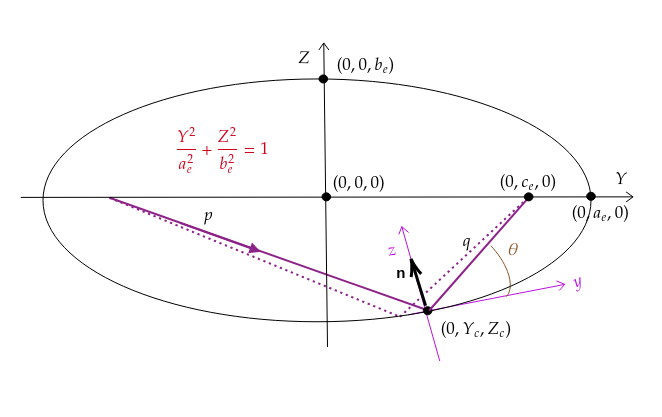}
    \caption{The ellipse in a centered reference system $(X,Y,Z)$. The local reference system is $(x,y,z)$.}
    \label{fig:ellipse}
\end{figure}

The meridional section in the plane $YZ$ has equation (Fig.~\ref{fig:ellipse}, Table~\ref{tab:centered_table}):
\begin{equation}
    \label{eq:ellipse}
    \frac{Y^2}{a_e^2}+\frac{Z^2}{b_e^2}=1
\end{equation}

The non-zero coefficients of the revolution ellipsoid are $c_{xx}~=~1/b_e^2$, $c_{yy}~=~1/a_e^2$, $c_{zz}~=~1/b_e^2$, $c_0~=~-1$. We have:
\begin{itemize}
    \item The design parameters are $p,~q, ~\theta$.
    \item The major axis is $a_e=(p+q)/2$; the minor axis is $b_e=\sqrt{p q}\sin\theta$, the foci are at at $(0,\pm c_e,0)$ with $c_e=\sqrt{a_e^2-b_e^2}$, and the eccentricity is $\epsilon=c_e/a_e$.
    \item  The mirror center is at: $X_c~=~0$, $Y_c=(p^2-q^2)/(4 c_e)~=~(p-q)/(2\epsilon)$, $Z_c=-b_e\sqrt{1-(Y_c/a_e)^2}=-p q \sin(2\theta)/(2c_e)=-b_e^2/( c_e \tan\theta)$ (see, e.g., Ref.~\cite{Goldberg-ellipse}).  
    \item The (non-normalized) normal at the mirror center is aligned with $\textbf{N}=(0,-2 Y_c / a_e^2, -2 Z_c / b_e^2) = (0,(q^2-p^2)/(2a_e^2 c_e), 2/(c_e \tan\theta))$.
    \item The unit normal vector is $\textbf{n} = (n_x,n_y,n_z) =\textbf{N}/|\textbf{N}|$.
    \item The angle from $\textbf{n}$ to $Z$ axis is $\Theta=\arcsin(n_y)$.
    \item We define $A=1/b_e^2$, $B=1/a_e^2$.
    The non-zero conic coefficients of the ellipsoid in the local mirror reference frame are
\begin{equation}
\begin{alignedat}{2}
    \label{eq:coeffs-ellipsoid}
    c'_{xx}&=A &&=b_e^{-2},\\
    c'_{yy}&= A n_y^2 + B n_z^2 &&=(n_y/b_e)^2 + (n_z/a_e)^2,\\
    c'_{zz}&= A n_z^2 + B n_y^2 &&=(n_z/b_e)^2 + (n_y/a_e)^2,\\
    c'_{yz}&= 2 n_y n_z (B-A) &&=2 n_y n_z (1/a_e^{2}-1/b_e^{2}),\\
    c'_{z} &= 2 (A n_z Z_c + B n_y Y_c) &&=2 (n_z Z_c/b_e^2 + n_y Y_c/a_e^2).
\end{alignedat}
\end{equation}
\end{itemize}

\subsection{Hyperboloid}

\begin{figure}
    \centering    
    \includegraphics[width=0.85\textwidth]{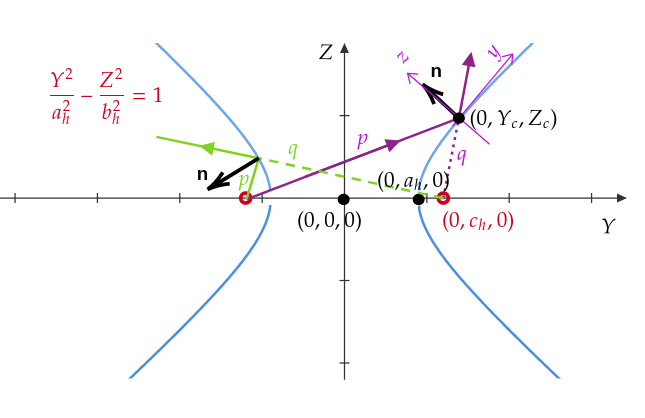}
    \caption{The hyperbola in a centered reference system $(X,Y,Z)$. The local reference system is $(x,y,z)$}
    \label{fig:hyperbola}
\end{figure}

The meridional section in the $YZ$ plane has equation (Fig.~\ref{fig:hyperbola}, Table~\ref{tab:centered_table}):
\begin{equation}
    \label{eq:hyperbola}
    \frac{Y^2}{a_h^2}-\frac{Z^2}{b_h^2}=1
\end{equation}

The non-zero coefficients of the revolution hyperboloid\footnote{As with the paraboloidal and ellipsoidal mirrors, We consider revolution surface around $Y$. This gives the \emph{two sheets hyperboloid} most relevant for optical elements. An alternative surface is produced by revolution around $X$ (\emph{one-sheet hyperboloid}).} are $c_{yy}=1/a_h^2, c_{zz}=c_{xx}=-1/b_h^2, c_0=-1$.
The major axis is $a_h=|p-q|/2$, the foci are at $(0,\pm c_h,0)$, with $c_h=(1/2)\sqrt{p^2+q^2-2 p q \cos(2 \theta)}$; the minor axis (imaginary) is $b_h=\sqrt{c_h^2-a_h^2}$.    
Two configurations are possible depending on the design parameters $p, q, \theta$:

i) Large $p$: $p>q$ (violet path in Fig.~\ref{fig:hyperbola}), where:
\begin{itemize}
    \item  The mirror center is at $X_c=0$, $Y_c=(p^2-q^2)/(4 c_h)$, $Z_c=b\sqrt{(Y_c/a_h)^2 - 1}$.  
    \item The (non-normalized) normal at the mirror center is aligned with $\textbf{N}=(0,-2 Y_c/a_h^2, 2 Z_c / b_h^2)$.
    \item The unit normal vector is $\textbf{n} = (n_x,n_y,n_z) = \textbf{N}/|\textbf{N}|$
    \item The angle from $\textbf{n}$ to $Z$ axis is $\Theta=\arcsin(n_y)$.
\end{itemize} 

ii) Large $q$: $q>p$ (green path in Fig.~\ref{fig:hyperbola}), where:
\begin{itemize}
    \item  The mirror center is at $X_c=0$, $Y_c=(p^2-q^2)/(4 c_h)$, $Z_c=b\sqrt{(Y_c/a_h)^2 - 1}$.  
    \item The (non-normalized) normal at the mirror center is $\textbf{N}=(0,2Y_c/a_h^2, -2 Z_c/b_h^2)$.
    \item The unit normal vector is $\textbf{n} = (n_x,n_y,n_z) = \textbf{N}/|\textbf{N}|$.
    \item The angle from $\textbf{n}$ to $Z$ axis is $\Theta=-\arccos(n_z)$.
\end{itemize} 

We define $A=-1/b_h^2$ and $B=1/a_h^2$. The non-zero conic coefficients of the hyperboloid in the local mirror reference frame are\footnote{the equations with parameters $A$ and $B$ are formally equal to those in Eqs.~(\ref{eq:coeffs-ellipsoid}) for the ellipsoid. However, $A<0$ for the hyperboloid and $A>0$ for the ellipsoid.}
\begin{equation}
\begin{alignedat}{2}
    \label{eq:coeffs-hyperboloid}
    c'_{xx}&=A &&=-b_h^2, \\
    c'_{yy}&= A n_{y}^2 + B n_{z}^2 &&=-(n_{y}/b_h)^2 + (n_{z}/a_h)^2, \\
    c'_{zz}&= A n_{z}^2 + B n_{y}^2 &&=-(n_{z}/b_h)^2 +  (n_{y}/a_h)^2, \\
    c'_{yz}&= 2 n_y n_z (B-A) &&= 2 n_y n_z (b_h^{-2}-a_h^{-2}), \\
    c'_{z}&= 2 (A n_{z} Z_c  + B n_{y} Y_c ) &&= 2 (- n_{z}  Z_c b_h^2  + n_{y} Y_c /a_h^2). 
\end{alignedat}
\end{equation}

\subsection{Degenerated surfaces}

A very important conic in optics is the sphere of radius $R$. It can be obtained directly from the ellipsoid Eqs.~(\ref{eq:coeffs-ellipsoid}) with $a=b=R, \textbf{n}=(0,0,1), \textbf{X}=(0,0,-R)$, therefore the non-zero coefficients are
\begin{equation}
    \label{eq:coeffs-sphere}
    \begin{split}
    c'_{xx}&=1, \\
    c'_{yy}&= 1, \\
    c'_{zz}&= 1, \\
    c'_{z}&= -2 R.
    \end{split}
\end{equation}

Another important surface is the plane, with equation $-z=0$, therefore the only non-zero coefficient is $c'_z=-1$ (the minus guarantees that the normal $\textbf{n}=-\bigtriangledown f$ is upwards).

The 2D curved surface of revolutionary conics (paraboloids, ellipsoids and hyperboloids) become cylinders (with parabolic, elliptical and hyperbolic sections, respectively) by making flat the surface along the sagittal direction ($x$). These surfaces that are flat in the sagittal direction are easily obtained from the revolution surfaces discussed before (paraboloids, ellipsoids and hyperboloids) by making zero the $c_i$ coefficients affecting the $x$ (i.e. $c_{xx}=c_{xy}=c_{xy}=c_{x}=0$). Similarly, a cylinder with axis parallel to the $y$ axis (flat in the meridional direction) has zero the coefficients affecting the $y$ (i.e. $c_{yy}=c_{xy}=c_{yz}=c_{y}=0$), but in these cases the cylinder section is circular.

\section{Discussion}

\subsection{Summary of equations for conics defined from design parameters parameters}

In section~\ref{sec:SectionConics} we obtained the coefficients of the conic in a non-centered system using roto-translation of the conic expressed in a centered system. The coefficients of the conic in the centered-system were calculated from the design parameters $p,q,\theta$. The rotation angle and translation vector are expressed as a function of the design parameters and the centered conic parameters. The position of the center and normal versor are summarized in Table~\ref{tab:centered_coordinates}, used for calculating the conic coefficients (Table~\ref{table:ccc-from-this-work}). 

\begin{table}[]
\centering
\begin{tabular}{cccccc}
    \hline
    & parabola  (focusing) & parabola  (collimating) & ellipse  &   hyperbola  $p>q$ & hyperbola  $p<q$ \\
\hline
$X_c$ & 0 & 0 & 0 & 0 & 0 \\
$Y_c$ & $-q\sin(2 \theta)$ & $p \sin(2\theta)$ & $\frac{p^2-q^2}{4c_e^2}$ & $\frac{p^2-q^2}{4c_e^2}$ & $\frac{p^2-q^2}{4c_e^2}$ \\
$Z_c$ & $q \cos^2\theta$ & $p \cos^2\theta$ & $-b_e\sqrt{1-\frac{Y_c^2}{a_e^2}}$ & $b_h\sqrt{\frac{Y_c^2}{a_h^2}-1}$ & $b_h\sqrt{\frac{Y_c^2}{a_h^2}-1}$  \\
\hline
$N_x$ & 0 & 0 & 0 & 0 & 0 \\
$N_y$ & $2 q \sin(2\theta)$ & $-2 p \sin(2\theta)$ & $\frac{-2Y_c}{a_e^2}=\frac{q^2-p^2}{2 a_e^2 c_e}$ & $\frac{-2Y_c}{a_h^2}$ & $\frac{2Y_c}{a_h^2}$ \\
$N_z$ & $4a_p$ & $4a_p$ & $\frac{-2Z_c}{b_e^2}=\frac{2}{c_e\tan\theta}$ & $\frac{2Z_c}{b_h^2}$ & $\frac{-2Z_c}{b_h^2}$ \\
\hline
$n_x$ & 0 & 0 & 0 & 0 & 0 \\
$n_y$ & $\cos\theta$ & $\sin\theta$ & $N_y/|\textbf{N}|$ & $N_y/|\textbf{N}|$ & $N_y/|\textbf{N}|$ \\
$n_z$ & $-\cos\theta$ & $\sin\theta$ & $N_z/|\textbf{N}|$ & $N_z/|\textbf{N}|$ & $N_z/|\textbf{N}|$ \\
\hline
$\Theta$ & $\frac{\pi}{2}-\theta$ & $-(\frac{\pi}{2}-\theta)$ & $\arcsin n_y$ & $\arcsin n_y$ & $-\arccos n_z$ \\
\hline
\end{tabular}
\caption{Coordinates of the center $(X_c,Y_c,Z_c)$, the normal $\textbf{N}=(N_x, N_y, N_z)$, the normalized normal $\textbf{n}=(n_x,n_y,n_z)$ with respect to the centered system for the parabola, ellipse and hyperbola. The angle $\Theta$ (from $\textbf{n}$ to the $Z$ axis) is also reported. 
}
\label{tab:centered_coordinates}
\end{table}

\begin{table}[H]
\centering
\begin{tabular}{lccc}
 Coefficient    & paraboloid & ellipsoid  & hyperboloid     \\
\hline
$c_{xx}$      & 1      &  $1$                        & $1$      \\ 
$c_{yy}$      & $n_z^2$  & $n_y^2 + \frac{b_e^2}{a_e^2} n_z^2$ & $n_{y}^2 - \frac{b_h^2}{a_h^2} n_{z}^2$      \\
$c_{zz}$      & $n_y^2$  & $n_z^2 + \frac{b_e^2}{a_e^2} n_y^2$ & $n_{z}^2 - \frac{b_h^2}{a_h^2} n_{y}^2$      \\
$c_{xy}$      & 0      & 0                            & 0      \\
$c_{yz}$      &$2 n_y n_z$ (or $-2 n_y n_z$) & $-2 n_y n_z (1-\frac{b_e^2}{a_e^2})$ & $-2 n_y n_z (1+\frac{b_h^2}{a_h^2})$     \\
$c_{xz}$      & 0      & 0                            & 0      \\
$c_{x}$       & 0      & 0                            & 0      \\
$c_{y}$       & 0      & 0                            & 0      \\
$c_{z}$       & $-4 \frac{a_p}{n_z}$& $2 (n_z Z_c + \frac{b_e^2}{a_e^2} n_y Y_c)$  & $2 (n_{z} Z_c  - \frac{b_h^2}{a_h^2} n_{y} Y_c)$      \\
$c_{0}$       & 0      & 0                            & 0      \\
\end{tabular}
\caption{Summary of conic coefficients as a function of the design parameters from section~\ref{sec:SectionConics} normalized to $c_{xx}=1$. In this table, $\textbf{n}=(0,n_y,n_z)$ is the normalized normal to the surface at the mirror pole (with coordinates $(0,Y_c,Z_c)$ in centered-coordinates). 
} 
\label{table:ccc-from-this-work}
\end{table}

In previous works, we studied parabolas \cite{Goldberg-parabola}, ellipses \cite{Goldberg-ellipse} and hyperbolas \cite{Goldberg-hyperbola} in non-centered systems, and deduced the equation of the optical surface from first principles. We applied Fermat principle to get the equation of the paraboloid, and ellipsoid, and geometrical principles for the hyperboloid. 
From these works, we summarize in Table~\ref{table:ccc-from-factory} the conic coefficients as a function of the design parameters for the paraboloid, ellipsoid, and hyperboloid surfaces.


\begin{table}[H]
\centering
\begin{tabular}{lccc}      
 Coefficient    & paraboloid  & ellipsoid  & hyperboloid     \\
\hline
$c_{xx}$      & 1      &  1            & 1      \\ 
$c_{yy}$      & $s^2$  & $s^2$         & $s^2$      \\
$c_{zz}$      & $c^2$  & $1-\left( s \frac{p-q}{p+q} \right) ^2$  & $1-\left( s \frac{p+q}{p-q} \right) ^2$       \\
$c_{xy}$      & 0             & 0      & 0      \\
$c_{yz}$      & $2 c s$ (or $-2c s$)   & $-2 s c \frac{q-p}{q+p}$    & $-2 s c \frac{q+p}{q-p}$     \\
$c_{xz}$      & 0             & 0      & 0      \\
$c_{x}$      & 0      & 0      & 0      \\
$c_{y}$      & 0      & 0      & 0      \\
$c_{z}$      & $-4 s q ~(\text{or} -4 s p)$      & $-4 s \frac{pq}{q+p}$     & $-4 s \frac{pq}{q-p}$      \\
$c_{0}$      & 0      & 0      & 0      \\
\end{tabular}
\caption{Conic coefficients as a function of the design parameters from Refs. \cite{Goldberg-parabola,Goldberg-ellipse,Goldberg-hyperbola}. $c=\cos\theta$, and $s=\sin\theta$. 
For the focusing paraboloid ($p=\infty$), use the first option; for the collimating paraboloid ($q=\infty$),  use the option in brackets).
}
\label{table:ccc-from-factory}
\end{table}

For consistency, we have checked that the coefficients calculated with the expressions from Tables~\ref{table:ccc-from-this-work} and ~\ref{table:ccc-from-factory} give the same results. 

\subsection{Numerical calculations.}
The Eqs.~(\ref{eq:transformedcoefficients}) is a general procedure very easy to implement on a computer. An example of calculation of the local coefficients of an ellipsoidal mirror is in Fig.~\ref{fig:pythoncode}.
\begin{figure}[H]
\begin{tiny}
    \input{pythoncode.tex}
\end{tiny}
    \caption{Example python code to calculate the conic coefficients of an ellipsoid mirror in the centered system, and transform them to the local reference system. The function {\tt rotate\_and\_translate\_coefficients} implements Eqs.~(\ref{eq:transformedcoefficients}). The main code implements some of the the Eqs. in sections~\ref{sec:conic-coefficients} and  \ref{sec:ellipse}.}
    \label{fig:pythoncode}
\end{figure}

\subsection{Misalignments and tolerances}
\begin{figure}
\centering
\includegraphics[width=0.85\textwidth,trim=0cm 0cm 0cm 0cm,clip=true]
{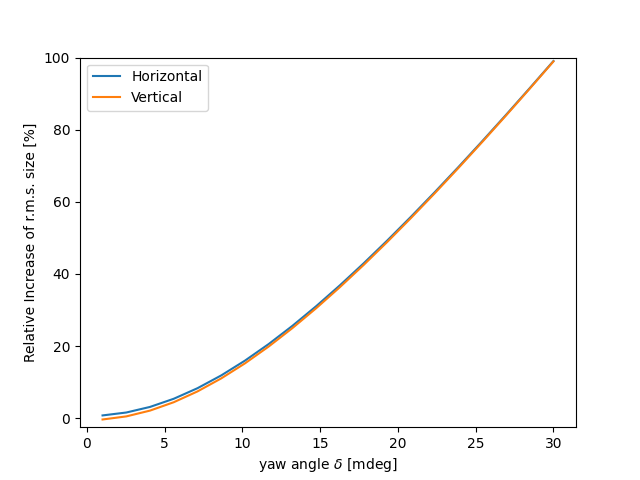}   
    \caption{Ray tracing calculation of the image size (relative increment with respect to the ideal size of \SI{0.1}{\micro\meter}) the image as a function of the misalignment in yaw angle $\delta$. For each value of $\delta$ a simulation is done using the transformed conic coefficients (see text).  Calculations are done in the ray tracing program SHADOW \cite{shadow}.
    }
\label{fig:tolerances}
\end{figure}

The general transformation of the conic coefficients in Eq.~(\ref{eq:transformedcoefficients}) can be used for studying possible misalignments in an optical system. A maximum accepted misalignment can be considered as a tolerance. We want to study, for example, the tolerances of an x-ray ellipsoidal mirror. We set design parameters to $p$=\SI{30}{\meter}, $q$=\SI{3}{\meter} and $\theta$=\SI{1}{\deg}. The conic coefficients (Table~\ref{table:ccc-from-this-work}) are [36.4793, 0.0111111, 36.4719, 0.0, 1.04164, 0.0, 0.0, 0.0, -6.9453, 0.0]. Suppose we want to study the tolerances on the yaw angle $\delta$, a rotation around the $z$ axis of matrix:
    \begin{equation}\label{eq:rotY}
    R_{M,y}(\delta) = \begin{bmatrix} 
    \cos\delta & 0 & \sin\delta \\
    0 & 1 & 0 \\
    -\sin\delta & 0 & \cos\delta
    \end{bmatrix}.
    \end{equation}
The translation $\textbf{t}$ is zero. A ray tracing simulation is designed using a source with Gaussian-shaped beam size of \SI{1}{\micro\meter} root mean square (r.m.s) and a divergence \SI{10}{\micro\radian} r.m.s, thus illuminating about \SI{170}{\milli\meter} of the mirror footprint. A ray tracing of the optical system is done for different values of $\delta$, and the r.m.s relative size increase of the image is displayed in Fig.~\ref{fig:tolerances} for the $x$ (horizontal) and $z$ (vertical) directions. If we accept an increase in relative size of 10\%, the plot indicates a maximum tolerance in the yaw angle of about \SI{16}{\milli\deg}.

\section{Summary and conclusions}

We have revisited the 3D conics used as optical surfaces in optics in general non-centered reference systems, like grazing optics. We deduced how these coefficients transform for a generic roto-translation. We then deduced the 10 coefficients as a function of the design parameters ($p,q,\theta$) for paraboloids, ellipsoids and hyperboloids. We consider spheres, planes, and a family of cylinders as 
particular degenerated cases. We illustrate the implementation of these coefficients and their transforms in a computed code in the python programming language. These results are of great utility for ray tracing simulations.  

While the descriptions can be cumbersome, once understood in context, these optical systems are both efficient and mathematically elegant. Stemming from an understanding of these foundational shapes, or abandoning them entirely with freeform elements, innovation and the increasing demands of scientific applications will dictate the direction of future designs.

\begin{backmatter}
\bmsection{Funding}
This work is supported by the Director, Office of Science, Office of Basic Energy Sciences of the U.S. Department of Energy (DE-AC02-05CH11231).








\bmsection{Data Availability Statement}
This work did not use any experimental data. The scripts corresponding to Figs.~\ref{fig:pythoncode} and \ref{fig:tolerances} other related software are available at the URL \href{https://github.com/srio/paper-conics-resources}{https://github.com/srio/paper-conics-resources}.

\end{backmatter}

\bibliography{sample}






\end{document}